\documentclass[12pt,letterpaper,english,aps,superscriptaddress,groupedaddress,showpacs,showkeys]{revtex4}
\usepackage[T1]{fontenc}
\usepackage[latin9]{inputenc}
\usepackage{babel}
\usepackage{amsmath}
\usepackage{amssymb}
\usepackage{esint}
\usepackage[unicode=true]
 {hyperref}
\usepackage{breakurl}

\makeatletter

\@ifundefined{textcolor}{}
{%
 \definecolor{BLACK}{gray}{0}
 \definecolor{WHITE}{gray}{1}
 \definecolor{RED}{rgb}{1,0,0}
 \definecolor{GREEN}{rgb}{0,1,0}
 \definecolor{BLUE}{rgb}{0,0,1}
 \definecolor{CYAN}{cmyk}{1,0,0,0}
 \definecolor{MAGENTA}{cmyk}{0,1,0,0}
 \definecolor{YELLOW}{cmyk}{0,0,1,0}
}

\usepackage{dcolumn}
\usepackage{bm}
\usepackage{amsfonts}

\makeatother

\begin{document}

\title{On the mean value of the force operator for 1D particles in the step
potential}

\author{Salvatore De Vincenzo}

\homepage{https://orcid.org/0000-0002-5009-053X}

\email{[salvatore.devincenzo@ucv.ve]}

\affiliation{Escuela de F\'{\i}sica, Facultad de Ciencias, Universidad Central de Venezuela,
A.P. 47145, Caracas 1041-A, Venezuela.}

\date{January 18, 2021}
\begin{abstract}
\noindent \textbf{Abstract} In the one-dimensional Klein-Fock-Gordon
theory, the probability density is a discontinuous function at the
point where the step potential is discontinuous. Thus, the mean value
of the external classical force operator cannot be calculated from
the corresponding formula of the mean value. To resolve this issue,
we obtain this quantity directly from the Klein-Fock-Gordon equation
in Hamiltonian form, or the Feshbach-Villars wave equation. Not without
surprise, the result obtained is not proportional to the average of
the discontinuity of the probability density but to the size of the
discontinuity. In contrast, in the one-dimensional Schr\"{o}dinger and
Dirac theories this quantity is proportional to the value that the
respective probability density takes at the point where the step potential
is discontinuous. We examine these issues in detail in this paper.
The presentation is suitable for the advanced undergraduate level.
\end{abstract}



\maketitle
\section{Introduction}

\noindent Let us consider a one-dimensional quantum particle in the
external finite step electric potential, i.e., 
\begin{equation}
\phi(x)=V_{0}\,\Theta(x),
\end{equation}
where $x\in\mathbb{R}$, $V_{0}=\mathrm{const}$, and $\Theta(x)$
is the Heaviside step function ($\Theta(x>0)=1$ and $\Theta(x<0)=0$).
Let us suppose that the wave function associated with the particle,
$\Psi=\Psi(x,t)$, can be normalizable, that is, $\Psi(x\rightarrow\pm\infty,t)=0$.
In fact, we will suppose that the norm of $\Psi$ is equal to one,
i.e., $\left\Vert \Psi\right\Vert ^{2}\,\equiv\langle\Psi,\Psi\rangle=1$,
where $\langle\;,\;\rangle$ denotes a specific scalar product (as
we will see below). Note: in the Klein-Fock-Gordon case, this norm
could also be equal to $-1$ (see, for example, Ref. \cite{RefA}). 

Now, let us write the one-dimensional Schr\"{o}dinger, Klein-Fock-Gordon,
and Dirac wave equations, in Hamiltonian form:
\begin{equation}
\mathrm{i}\hbar\frac{\partial}{\partial t}\Psi=\hat{\mathrm{h}}_{\mathrm{A}}\Psi\,,\quad\mathrm{A}=\mathrm{S},\mathrm{KFG},\mathrm{D}\,,
\end{equation}
where 
\begin{equation}
\hat{\mathrm{h}}_{\mathrm{S}}=-\frac{\hbar^{2}}{2\mathrm{m}}\frac{\partial^{2}}{\partial x^{2}}+\phi(x)
\end{equation}
is the Schr\"{o}dinger Hamiltonian operator; 
\begin{equation}
\hat{\mathrm{h}}_{\mathrm{KFG}}=-\frac{\hbar^{2}}{2\mathrm{m}}\left(\hat{\tau}_{3}+\mathrm{i}\hat{\tau}_{2}\right)\frac{\partial^{2}}{\partial x^{2}}+\mathrm{m}c^{2}\hat{\tau}_{3}+\phi(x),
\end{equation}
is --let us say-- the Klein-Fock-Gordon Hamiltonian operator (where
$\hat{\tau}_{3}=\left[\begin{array}{cc}
1 & 0\\
0 & -1
\end{array}\right]$ and $\hat{\tau}_{2}=\left[\begin{array}{cc}
0 & -\mathrm{i}\\
\mathrm{i} & 0
\end{array}\right]$) and the respective wave equation is called the Feshbach-Villars
equation \cite{RefA,RefB,RefC,RefD}, and 
\begin{equation}
\hat{\mathrm{h}}_{\mathrm{D}}=-\mathrm{i}\hbar c\,\hat{\alpha}\frac{\partial}{\partial x}+\mathrm{m}c^{2}\hat{\beta}+\phi(x),
\end{equation}
is the Dirac Hamiltonian operator (where $\hat{\alpha}$ and $\hat{\beta}$
are the $2\times2$ Dirac matrices, which satisfy the relations $\hat{\alpha}^{2}=\hat{\beta}^{2}=\hat{1}$
and $\hat{\alpha}\hat{\beta}+\hat{\beta}\hat{\alpha}=\hat{0}$).

Certainly, the operator $\hat{\mathrm{h}}_{\mathrm{S}}$ acts on one-component
wave functions, and the scalar product for these wave functions is
the usual one, i.e., $\langle\Psi,\Phi\rangle_{\mathrm{S}}=\int_{\mathbb{R}}\mathrm{d}x\,\Psi^{*}\Phi$
(the asterisk $^{*}$ denotes the complex conjugate, as usual). Also,
$\left\Vert \Psi\right\Vert _{\mathrm{S}}^{2}\,\equiv\langle\Psi,\Psi\rangle_{\mathrm{S}}=\int_{\mathbb{R}}\mathrm{d}x\,\varrho_{\mathrm{S}}$,
where $\varrho_{\mathrm{S}}=\varrho_{\mathrm{S}}(x,t)=|\Psi(x,t)|^{2}$
is the Schr\"{o}dinger probability density. On the other hand, the operator
$\hat{\mathrm{h}}_{\mathrm{\mathrm{KFG}}}$ acts on two-component
column vectors of the form $\Psi=\left[\,\varphi\;\,\chi\,\right]^{\mathrm{T}}$
(the symbol $^{\mathrm{T}}$ represents the transpose of a matrix).
In this case, the scalar product must be defined as $\langle\Psi,\Phi\rangle_{\mathrm{KFG}}=\int_{\mathbb{R}}\mathrm{d}x\,\Psi^{\dagger}\hat{\tau}_{3}\Phi$
(the symbol $^{\dagger}$ denotes the Hermitian conjugate, or the
adjoint, of a matrix and an operator) \cite{RefA,RefB,RefC,RefD}.
Also, $\left\Vert \Psi\right\Vert _{\mathrm{\mathrm{KFG}}}^{2}\,\equiv\langle\Psi,\Psi\rangle_{\mathrm{KFG}}=\int_{\mathbb{R}}\mathrm{d}x\,\varrho_{\mathrm{KFG}}$,
where $\varrho_{\mathrm{KFG}}=\varrho_{\mathrm{KFG}}(x,t)=\Psi^{\dagger}(x,t)\hat{\tau}_{3}\Psi(x,t)=|\varphi(x,t)|^{2}-|\chi(x,t)|^{2}$
is --let us say-- the Klein-Fock-Gordon probability density, although
we know that calling this quantity probability density is not absolutely
correct (this is because it is not positive definite), see, for example,
Refs. \cite{RefA,RefB,RefC,RefD}. Likewise, the operator $\hat{\mathrm{h}}_{\mathrm{D}}$
acts on two-component wave functions of the form $\Psi=\left[\,\psi_{1}\;\,\psi_{2}\,\right]^{\mathrm{T}}$,
and the scalar product for these wave functions is given by $\langle\Psi,\Phi\rangle_{\mathrm{D}}=\int_{\mathbb{R}}\mathrm{d}x\,\Psi^{\dagger}\Phi$.
Also, $\left\Vert \Psi\right\Vert _{\mathrm{D}}^{2}\,\equiv\langle\Psi,\Psi\rangle_{\mathrm{D}}=\int_{\mathbb{R}}\mathrm{d}x\,\varrho_{\mathrm{D}}$,
where $\varrho_{\mathrm{D}}=\varrho_{\mathrm{D}}(x,t)=\Psi^{\dagger}(x,t)\Psi(x,t)=|\psi_{1}(x,t)|^{2}+|\psi_{2}(x,t)|^{2}$
is the Dirac probability density. Finally, as is known, Hamiltonians
$\hat{\mathrm{h}}_{\mathrm{S}}$ and $\hat{\mathrm{h}}_{\mathrm{D}}$
are Hermitian, or formally self-adjoint, differential operators, i.e.,
$\hat{\mathrm{h}}_{\mathrm{S}}=\hat{\mathrm{h}}_{\mathrm{S}}^{\dagger}$
and $\hat{\mathrm{h}}_{\mathrm{D}}=\hat{\mathrm{h}}_{\mathrm{D}}^{\dagger}$,
thus, $\langle\Psi,\hat{\mathrm{h}}_{\mathrm{A}}\Phi\rangle_{\mathrm{A}}=\langle\hat{\mathrm{h}}_{\mathrm{A}}\Psi,\Phi\rangle_{\mathrm{A}}$
($\mathrm{A}=\mathrm{S},\mathrm{D}$). Because $\hat{\mathrm{h}}_{\mathrm{KFG}}=\hat{\tau}_{3}\hat{\mathrm{h}}_{\mathrm{KFG}}^{\dagger}\hat{\tau}_{3}$,
the Hamiltonian $\hat{\mathrm{h}}_{\mathrm{KFG}}$ is called pseudo-Hermitian,
or formally pseudo-self-adjoint, i.e., $\langle\Psi,\hat{\mathrm{h}}_{\mathrm{KFG}}\Phi\rangle_{\mathrm{KFG}}=\langle\hat{\mathrm{h}}_{\mathrm{KFG}}\Psi,\Phi\rangle_{\mathrm{KFG}}$. 

Let us think about obtaining the average value (or the mean value)
of the external classical force operator,
\begin{equation}
\hat{f}=-\mathrm{d}\phi(x)/\mathrm{d}x=-V_{0}\,\delta(x),
\end{equation}
directly from each wave equation ($\delta(x)=\mathrm{d}\Theta(x)/\mathrm{d}x$
is the Dirac delta function), i.e., let us think about obtaining the
following three quantities: $\langle\hat{f}\rangle_{\mathrm{A}}=\int_{\mathbb{R}}\mathrm{d}x\,\hat{f}\varrho_{\mathrm{A}}$,
with $\mathrm{A}=\mathrm{S},\mathrm{KFG},\mathrm{D}$, directly from
the respective wave equation (even knowing that the calculation of
these quantities from the previous formula seems absolutely immediate).
Certainly, the procedure to obtain each average value depends on the
respective equation, although all three are similar procedures. For
example, if the Schr\"{o}dinger equation is considered, the procedure
is as follows: (a) multiply the wave equation for $\Psi$ by $\partial\Psi^{*}/\partial x$
and the equation for $\Psi^{*}$ by $\partial\Psi/\partial x$, and
then sum the two resulting equations; (b) integrate each term of the
result obtained in (a) around $x=0$. Then, we obtain the following
result:
\begin{equation}
0=-\frac{\hbar^{2}}{2\mathrm{m}}\left.\left[\,|\Psi_{x}(x,t)|^{2}\,\right]\:\right|_{0-}^{0+}+\left.\left[\,\phi(x)\varrho_{\mathrm{S}}(x,t)\,\right]\:\right|_{0-}^{0+}+\langle\hat{f}\rangle_{\mathrm{S}},
\end{equation}
where we use hereafter the notation $\left.\left[\, g\,\right]\:\right|_{0-}^{0+}\equiv g(0+,t)-g(0-,t)$
with $0\pm\equiv\underset{\epsilon\rightarrow0}{\lim}\,(0\pm\epsilon)$,
and also $\Psi_{x}\equiv\partial\Psi/\partial x$. For the step potential
given in Eq. (1), we have that $\Psi(0+,t)=\Psi(0-,t)\equiv\Psi(0,t)$
and $\Psi_{x}(0+,t)=\Psi_{x}(0-,t)\equiv\Psi_{x}(0,t)$, i.e., $\Psi(x,t)$
and $\Psi_{x}(x,t)$, and therefore, $\varrho_{\mathrm{S}}(x,t)$,
are continuous functions at $x=0$. We finally obtain the result
\begin{equation}
\langle\hat{f}\rangle_{\mathrm{S}}=-V_{0}\,\varrho_{\mathrm{S}}(0,t),
\end{equation}
which is also obtained immediately from the formula to calculate the
average value of the operator $\hat{f}$ in the Schr\"{o}dinger case,
namely, $\langle\hat{f}\rangle_{\mathrm{S}}=-V_{0}\int_{\mathbb{R}}\mathrm{d}x\,\delta(x)\varrho_{\mathrm{S}}(x,t)=-V_{0}\,\varrho_{\mathrm{S}}(0,t)$.
In the latter result we used Eq. (6) and the defining property of
the Dirac delta function, which is valid for continuous functions
at the single point where the Dirac delta is infinite. It is worth
noting that, the relation we have written in Eq. (7) gives us the
average value of the external classical force operator even when $V_{0}\rightarrow\infty$
(see appendix A).

Similarly, if the Dirac equation is considered, the procedure is as
follows: (a) multiply (properly) the wave equation for $\Psi$ by
$\partial\Psi^{\dagger}/\partial x$ and the equation for $\Psi^{\dagger}$
by $\partial\Psi/\partial x$, and then sum the two resulting equations;
(b) integrate each term of the resulting expression obtained in (a)
over $x=0$. Then, we obtain the following result:
\begin{equation}
0=\mathrm{m}c^{2}\left.\left[\,\Psi^{\dagger}(x,t)\hat{\beta}\Psi(x,t)\,\right]\:\right|_{0-}^{0+}+\left.\left[\,\phi(x)\varrho_{\mathrm{D}}(x,t)\,\right]\:\right|_{0-}^{0+}+\langle\hat{f}\rangle_{\mathrm{D}}.
\end{equation}
For the potential step in Eq. (1), we have just that $\Psi(0+,t)=\Psi(0-,t)\equiv\Psi(0,t)$,
thus, $\Psi^{\dagger}(x,t)\hat{\beta}\Psi(x,t)$ and $\varrho_{\mathrm{D}}(x,t)$
are also continuous functions at $x=0$, and we obtain the result
\begin{equation}
\langle\hat{f}\rangle_{\mathrm{D}}=-V_{0}\,\varrho_{\mathrm{D}}(0,t).
\end{equation}
Certainly, from the formula to calculate the average value of the
operator $\hat{f}$ in the Dirac case, the result in Eq. (10) can
also be obtained immediately, namely, $\langle\hat{f}\rangle_{\mathrm{D}}=-V_{0}\int_{\mathbb{R}}\mathrm{d}x\,\delta(x)\varrho_{\mathrm{D}}(x,t)=-V_{0}\,\varrho_{\mathrm{D}}(0,t)$.
Again, in the latter result we used Eq. (6) and the key property of
the Dirac delta function (remember that the latter property only works
for functions that are continuous at $x=0$). Incidentally, in Ref.
\cite{RefE}, the Ehrenfest theorem for the one-dimensional Dirac
particle in the finite step potential was studied, and it was demonstrated
that the formal time derivative of the mean value of the one-dimensional
(Dirac) momentum operator, $\hat{\mathrm{p}}=-\mathrm{i}\hbar\hat{1}\partial/\partial x$,
is precisely $\langle\hat{f}\rangle_{\mathrm{D}}$. The result given
in Eq. (9) was also obtained in the same reference. 

\section{The mean value of $\hat{f}$ in the Klein-Fock-Gordon case}

\noindent Now, we consider the Hamiltonian form of the Klein-Fock-Gordon
equation in one spatial dimension, or the one-dimensional first order
in time Klein-Fock-Gordon equation, or the one-dimensional Feshbach-Villars
wave equation. The procedure to obtain the average value of the operator
$\hat{f}$ in Eq. (6) from the latter equation is as follows: (a)
multiply (properly) the equation for $\Psi$ by $(\partial\Psi^{\dagger}/\partial x)\hat{\tau}_{3}$
and the equation for $\Psi^{\dagger}$ by $\hat{\tau}_{3}\,\partial\Psi/\partial x$,
and then sum the two resulting equations; (b) integrate each term
of the resulting expression obtained in (a) across $x=0$. Thus, we
obtain 
\[
0=-\frac{\hbar^{2}}{2\mathrm{m}}\left.\left[\,\Psi_{x}^{\dagger}(x,t)(\hat{1}+\hat{\tau}_{1})\Psi_{x}(x,t)\,\right]\right|_{0-}^{0+}+\mathrm{m}c^{2}\left.\left[\,\Psi^{\dagger}(x,t)\Psi(x,t)\,\right]\right|_{0-}^{0+}
\]
\begin{equation}
+\left.\left[\,\phi(x)\varrho_{\mathrm{KFG}}(x,t)\,\right]\:\right|_{0-}^{0+}+\langle\hat{f}\rangle_{\mathrm{KGF}}
\end{equation}
(where $\hat{\tau}_{1}=\left[\begin{array}{cc}
0 & 1\\
1 & 0
\end{array}\right]$). For a discontinuous potential at $x=0$, such as that given in
Eq. (1), we have to impose the following boundary conditions: $\psi(0+,t)=\psi(0-,t)\equiv\psi(0,t)$
and $\psi_{x}(0+,t)=\psi_{x}(0-,t)\equiv\psi_{x}(0,t)$, where $\psi=\psi(x,t)$
is the solution of the one-dimensional Klein-Fock-Gordon wave equation
in its standard form, or the second order in time Klein-Fock-Gordon
equation in one spatial dimension \cite{RefF,RefG,RefH,RefI}, namely,
\begin{equation}
\left[\,\mathrm{i}\hbar\frac{\partial}{\partial t}-\phi(x)\right]^{2}\psi=\left[-\hbar^{2}c^{2}\frac{\partial^{2}}{\partial x^{2}}+(\mathrm{m}c^{2})^{2}\right]\psi
\end{equation}
(in fact, the latter equation involves second as well as first time
derivatives). The continuity of $\psi$ and $\psi_{x}$ at $x=0$
implies that the Klein-Fock-Gordon probability current density is
also continuous there (see, for example, Refs. \cite{RefC,RefD}).
In (3+1) dimensions, this result is also true if there is no potential
vector, and also if it is different from zero but continuous at $x=0$.
Because the current and probability densities satisfy the usual continuity
equation, this last result and the continuity of the probability current
density at $x=0$ imply that the integral of the probability density
over the small interval around the point $x=0$ does not change in
time (what is an expected result). Incidentally, in Ref. \cite{RefJ},
among other things, it was verified that the stationary solutions
of Eq. (12) with the step potential given in Eq. (1), and also its
first spatial derivatives, must be continuous at $x=0$. In fact,
in that reference, the stationary solutions of the Klein-Fock-Gordon
equation with an smooth potential that tends to the step potential
after taking a limit were first obtained, then, by taking the same
limit on these solutions, the result in question was verified. Also,
the relation between $\psi$ (a one-component wave function) and $\Psi$
(a two-component column vector), is given by
\begin{equation}
\Psi=\left[\begin{array}{c}
\varphi\\
\chi
\end{array}\right]=\frac{1}{2}\left[\begin{array}{c}
\psi+\frac{1}{\mathrm{m}c^{2}}\left(\mathrm{i}\hbar\frac{\partial}{\partial t}-\phi\right)\psi\\
\psi-\frac{1}{\mathrm{m}c^{2}}\left(\mathrm{i}\hbar\frac{\partial}{\partial t}-\phi\right)\psi
\end{array}\right],
\end{equation}
and from the latter expression we obtain 
\begin{equation}
\psi=\varphi+\chi\,,\quad\mathrm{and}\quad\frac{1}{\mathrm{m}c^{2}}\left(\mathrm{i}\hbar\frac{\partial}{\partial t}-\phi\right)\psi=\varphi-\chi.
\end{equation}
These relations can be used to write $\varrho_{\mathrm{KFG}}=\Psi^{\dagger}\hat{\tau}_{3}\Psi=|\varphi|^{2}-|\chi|^{2}$
in terms of $\psi$. In effect, the following well-known result is
obtained:
\begin{equation}
\varrho_{\mathrm{KFG}}=\frac{\mathrm{i}\hbar}{2\mathrm{m}c^{2}}\left(\psi^{*}\dot{\psi}-\psi\dot{\psi}^{*}\right)-\frac{\phi}{\mathrm{m}c^{2}}\psi^{*}\psi,
\end{equation}
where $\dot{\psi}\equiv\partial\psi/\partial t$ (see, for example,
Ref. \cite{RefC}). 

Because the potential in Eq. (1) vanishes for $x<0$, and has the
value $V_{0}$ for $x>0$, we can write the following relations from
Eq. (13):
\begin{equation}
\Psi=\frac{1}{2}\left[\begin{array}{c}
\psi+\frac{\mathrm{i}\hbar}{\mathrm{m}c^{2}}\dot{\psi}\\
\psi-\frac{\mathrm{i}\hbar}{\mathrm{m}c^{2}}\dot{\psi}
\end{array}\right]\quad\left(\:\Rightarrow\:\Psi_{x}=\frac{1}{2}\left[\begin{array}{c}
\psi_{x}+\frac{\mathrm{i}\hbar}{\mathrm{m}c^{2}}\dot{\psi}_{x}\\
\psi_{x}-\frac{\mathrm{i}\hbar}{\mathrm{m}c^{2}}\dot{\psi}_{x}
\end{array}\right]\:\right),
\end{equation}
for $x<0$, and 
\begin{equation}
\Psi=\frac{1}{2}\left[\begin{array}{c}
\psi+\frac{\mathrm{i}\hbar}{\mathrm{m}c^{2}}\dot{\psi}-\frac{V_{0}}{\mathrm{m}c^{2}}\psi\\
\psi-\frac{\mathrm{i}\hbar}{\mathrm{m}c^{2}}\dot{\psi}+\frac{V_{0}}{\mathrm{m}c^{2}}\psi
\end{array}\right]\quad\left(\:\Rightarrow\:\Psi_{x}=\frac{1}{2}\left[\begin{array}{c}
\psi_{x}+\frac{\mathrm{i}\hbar}{\mathrm{m}c^{2}}\dot{\psi}_{x}-\frac{V_{0}}{\mathrm{m}c^{2}}\psi_{x}\\
\psi_{x}-\frac{\mathrm{i}\hbar}{\mathrm{m}c^{2}}\dot{\psi}_{x}+\frac{V_{0}}{\mathrm{m}c^{2}}\psi_{x}
\end{array}\right]\:\right),
\end{equation}
for $x>0$ (we also have that $\Psi_{x}=\left[\,\varphi_{x}\;\,\chi_{x}\,\right]^{\mathrm{T}}$).
Thus, from Eqs. (16) and (17), and using the continuity of $\psi$
and $\psi_{x}$ at $x=0$, we obtain the following matricial boundary
conditions: 
\begin{equation}
\Psi(0+,t)=\Psi(0-,t)+\frac{1}{2}\frac{V_{0}}{\mathrm{m}c^{2}}\left[\begin{array}{c}
-1\\
+1
\end{array}\right]\psi(0,t)
\end{equation}
and 
\begin{equation}
\Psi_{x}(0+,t)=\Psi_{x}(0-,t)+\frac{1}{2}\frac{V_{0}}{\mathrm{m}c^{2}}\left[\begin{array}{c}
-1\\
+1
\end{array}\right]\psi_{x}(0,t).
\end{equation}
Thus, we have that the column vector $\Psi$ and its first spatial
derivative $\Psi_{x}$ are discontinuous functions at $x=0$. Likewise,
the Klein-Fock-Gordon probability density is also discontinuous at
$x=0$. Precisely, from Eq. (15) we obtain the following result pertaining
to this discontinuity: 
\begin{equation}
\varrho_{\mathrm{KFG}}(0+,t)=\varrho_{\mathrm{KFG}}(0-,t)-\frac{V_{0}}{\mathrm{m}c^{2}}\psi^{*}(0,t)\psi(0,t).
\end{equation}
Incidentally, the boundary conditions in Eqs. (18) and (19) can be
written in a way reminiscent of the periodic boundary condition (i.e.,
thinking in a real line with the origin excluded, or a hole, between
$x=0-$ and $x=0+$), namely,
\begin{equation}
(\hat{\tau}_{3}+\mathrm{i}\hat{\tau}_{2})\Psi(0+,t)=(\hat{\tau}_{3}+\mathrm{i}\hat{\tau}_{2})\Psi(0-,t),
\end{equation}
and 
\begin{equation}
(\hat{\tau}_{3}+\mathrm{i}\hat{\tau}_{2})\Psi_{x}(0+,t)=(\hat{\tau}_{3}+\mathrm{i}\hat{\tau}_{2})\Psi_{x}(0-,t),
\end{equation}
respectively; hence, $(\hat{\tau}_{3}+\mathrm{i}\hat{\tau}_{2})\Psi$
and $(\hat{\tau}_{3}+\mathrm{i}\hat{\tau}_{2})\Psi_{x}$ are continuous
functions at $x=0$. Clearly, this result is obtained by multiplying
Eqs. (18) and (19) from the left by the matrix $\hat{\tau}_{3}+\mathrm{i}\hat{\tau}_{2}$.
The latter two boundary conditions can also be obtained directly from
the one-dimensional Feshbach-Villars wave equation (see appendix B). 

Finally, to obtain $\langle\hat{f}\rangle_{\mathrm{KFG}}$ from the
formula (11), we must compute the three boundary terms that are present
there. Thus, we can report the following results: 
\begin{equation}
\left.\left[\,\phi(x)\varrho_{\mathrm{KFG}}(x,t)\,\right]\:\right|_{0-}^{0+}=V_{0}\,\varrho_{\mathrm{KFG}}(0+,t),
\end{equation}
which is obtained immediately. Additionally, 
\begin{equation}
\mathrm{m}c^{2}\left.\left[\,\Psi^{\dagger}(x,t)\Psi(x,t)\,\right]\right|_{0-}^{0+}=-\frac{V_{0}}{2}\left[\,\varrho_{\mathrm{KFG}}(0+,t)+\varrho_{\mathrm{KFG}}(0-,t)\,\right],
\end{equation}
where we have used the result given in Eq. (18), the relation on the
right side of Eq. (14), and Eq. (15), both evaluated at $x=0-$, and
Eq. (20). Likewise, 
\begin{equation}
-\frac{\hbar^{2}}{2\mathrm{m}}\left.\left[\,\Psi_{x}^{\dagger}(x,t)(\hat{1}+\hat{\tau}_{1})\Psi_{x}(x,t)\,\right]\right|_{0-}^{0+}=0,
\end{equation}
where we have used the result given in Eq. (19). Then, substituting
the three latter results into Eq. (11), we obtain the result
\begin{equation}
\langle\hat{f}\rangle_{\mathrm{KFG}}=-\frac{V_{0}}{2}\left[\,\varrho_{\mathrm{KFG}}(0+,t)-\varrho_{\mathrm{KFG}}(0-,t)\,\right].
\end{equation}
From the formula to calculate the average value of the operator $\hat{f}$
in the Klein-Fock-Gordon case, the result in Eq. (26) cannot be derived.
In effect, $\langle\hat{f}\rangle_{\mathrm{KFG}}=\int_{\mathbb{R}}\mathrm{d}x\,\hat{f}\varrho_{\mathrm{KFG}}(x,t)=-V_{0}\int_{\mathbb{R}}\mathrm{d}x\,\delta(x)\varrho_{\mathrm{KFG}}(x,t)$,
but $\varrho_{\mathrm{KFG}}(x,t)$ is not continuous at $x=0$ (see
Eq. (20)). Thus, following this last procedure we could not say anything
else, but, as we have seen, the first order in time Klein-Fock-Gordon
equation can determine precisely what is the value of the integral
$\int_{\mathbb{R}}\mathrm{d}x\,\delta\,\varrho_{\mathrm{KFG}}$. In
effect, we can write
\begin{equation}
\int_{\mathbb{R}}\mathrm{d}x\,\delta(x)\varrho_{\mathrm{KFG}}(x,t)=\frac{1}{2}\left[\,\varrho_{\mathrm{\mathrm{KFG}}}(0+,t)-\varrho_{\mathrm{\mathrm{KFG}}}(0-,t)\,\right].
\end{equation}
From Eq. (20) we can also write $\langle\hat{f}\rangle_{\mathrm{\mathrm{KFG}}}$
in terms of the one-component wave function $\psi$, namely, 
\begin{equation}
\langle\hat{f}\rangle_{\mathrm{\mathrm{KFG}}}=+\frac{V_{0}}{2}\frac{V_{0}}{\mathrm{m}c^{2}}\psi^{*}(0,t)\psi(0,t).
\end{equation}
It is worth mentioning that, the result in Eq. (26) confirms that
an integral like the one given in the left side of Eq. (27) is not
always equal to the average of the discontinuity of the function that
accompanies the Dirac delta in the integral (in this case, the function
$\varrho_{\mathrm{\mathrm{KFG}}}$), i.e., the right side of Eq. (27)
is not necessarily equal to $\tfrac{1}{2}[\,\varrho_{\mathrm{KFG}}(0+,t)+\varrho_{\mathrm{\mathrm{KFG}}}(0-,t)\,]$.
For other interesting examples in which similar situations arise,
see Refs. \cite{RefK,RefL,RefM}.

Let us finally develop the nonrelativistic approximation of the result
in Eq. (26) (or the result in Eq. (28)). In order to do this, we first
write the Klein-Fock-Gordon wave function in the form $\psi(x,t)=\Psi_{\mathrm{S}}(x,t)\exp(-\mathrm{i}\,\mathrm{m}c^{2}t/\hbar)$
(here, $\Psi_{\mathrm{S}}$ is the Schr\"{o}dinger wave function), and
therefore, $\dot{\psi}(x,t)=(-\mathrm{i}\,\mathrm{m}c^{2}/\hbar)\Psi_{\mathrm{S}}(x,t)\exp(-\mathrm{i}\,\mathrm{m}c^{2}t/\hbar)$
(in this limit, we have that $\left|\,\mathrm{i}\hbar\dot{\Psi}_{\mathrm{S}}\,\right|\ll\mathrm{m}c^{2}\left|\,\Psi_{\mathrm{S}}\,\right|$)
\cite{RefD}. Then, we note that, in the nonrelativistic limit, the
Klein-Fock-Gordon probability density given in Eq. (15) reduces to
\begin{equation}
\varrho_{\mathrm{KFG}}=\varrho_{\mathrm{S}}-\frac{\phi}{\mathrm{m}c^{2}}\varrho_{\mathrm{S}}
\end{equation}
(remember that $\varrho_{\mathrm{S}}=\Psi_{\mathrm{S}}^{*}\Psi_{\mathrm{S}}$).
Using this result, we can also write $\langle\hat{f}\rangle_{\mathrm{KFG}}$
in this limit (see Eq. (26)), namely,
\[
\langle\hat{f}\rangle_{\mathrm{KFG}}=-\frac{V_{0}}{2}\left[\varrho_{\mathrm{S}}(0,t)-\frac{\phi(0+)}{\mathrm{m}c^{2}}\varrho_{\mathrm{S}}(0,t)\right]+\frac{V_{0}}{2}\left[\varrho_{\mathrm{S}}(0,t)-\frac{\phi(0-)}{\mathrm{m}c^{2}}\varrho_{\mathrm{S}}(0,t)\right]
\]
\begin{equation}
=+\frac{V_{0}}{2}\frac{V_{0}}{\mathrm{m}c^{2}}\varrho_{\mathrm{S}}(0,t)
\end{equation}
(which can also be obtained directly from Eq. (28)). Likewise, using
the result in Eq. (29), we can write
\begin{equation}
\langle\hat{f}\rangle_{\mathrm{KFG}}=\int_{\mathbb{R}}\mathrm{d}x\,\hat{f}\varrho_{\mathrm{\mathrm{KFG}}}=\langle\hat{f}\rangle_{\mathrm{S}}+\frac{V_{0}}{\mathrm{m}c^{2}}\int_{\mathbb{R}}\mathrm{d}x\,\delta(x)\phi(x)\varrho_{\mathrm{S}}(x,t),
\end{equation}
but the latter integral cannot be precisely evaluated because $\phi(x)$
is not continuous at $x=0$. Despite that, we decide to define the
value of that integral as $\phi(0)\varrho_{\mathrm{S}}(0,t)$, and
therefore, the result in Eq. (31) is as follows:
\begin{equation}
\langle\hat{f}\rangle_{\mathrm{\mathrm{KFG}}}=\int_{\mathbb{R}}\mathrm{d}x\,\hat{f}\varrho_{\mathrm{KFG}}=\langle\hat{f}\rangle_{\mathrm{S}}+\frac{V_{0}}{\mathrm{m}c^{2}}\phi(0)\varrho_{\mathrm{S}}(0,t).
\end{equation}
Finally, by equating the results given in Eqs. (30) and (32), we obtain
\begin{equation}
\langle\hat{f}\rangle_{\mathrm{S}}=-\left[\phi(0)-\frac{V_{0}}{2}\right]\frac{V_{0}}{\mathrm{m}c^{2}}\varrho_{\mathrm{S}}(0,t).
\end{equation}
Certainly, we already know what this last result should be, namely,
$\langle\hat{f}\rangle_{\mathrm{S}}=-V_{0}\,\varrho_{\mathrm{S}}(0,t)$
(Eq. (8)). Thus, in the nonrelativistic limit, $\tfrac{V_{0}}{2}+\mathrm{m}c^{2}$
should tend to $\phi(0)$. In any case, the latter result can be obtained
as follows. First, the one-dimensional Klein-Fock-Gordon wave equation
in Hamiltonian form is fully equivalent to the following system of
coupled differential equations:
\begin{equation}
\mathrm{i}\hbar\frac{\partial}{\partial t}\varphi=-\frac{\hbar^{2}}{2\mathrm{m}}\frac{\partial^{2}}{\partial x^{2}}\left(\varphi+\chi\right)+\phi\varphi+\mathrm{m}c^{2}\varphi\,,\quad\mathrm{i}\hbar\frac{\partial}{\partial t}\chi=+\frac{\hbar^{2}}{2\mathrm{m}}\frac{\partial^{2}}{\partial x^{2}}\left(\varphi+\chi\right)+\phi\chi-\mathrm{m}c^{2}\chi,
\end{equation}
where the top and bottom components of the column vector $\Psi$,
$\varphi$ and $\chi$, are given in terms of $\psi$ in Eq. (13).
Then, if we consider the ansatz $\psi=\Psi_{\mathrm{S}}\exp(-\mathrm{i}\,\mathrm{m}c^{2}t/\hbar)$,
and therefore, $\dot{\psi}=(-\mathrm{i}\,\mathrm{m}c^{2}/\hbar)\psi$
(the latter in the nonrelativistic limit), we obtain, in this approximation,
$\varphi=\left(1-\tfrac{\phi}{2\mathrm{m}c^{2}}\right)\psi$ (see
Eq. (13)). We also have that $\varphi+\chi=\psi$ (see Eq. (14)).
Thus, for weak external fields (to lowest order), $\varphi\approx\psi$
satisfies the nonrelativistic Schr\"{o}dinger equation but with the replacement
$\phi+\mathrm{m}c^{2}\rightarrow\phi$ (see the first equation in
Eq. (34)), and therefore, $\phi(0)+\mathrm{m}c^{2}\rightarrow\phi(0)$,
and also $\tfrac{V_{0}}{2}+\mathrm{m}c^{2}\rightarrow\phi(0)(\equiv\tfrac{V_{0}}{2})$
(in the last relation, we set the value $\phi\equiv\tfrac{V_{0}}{2}$
at $x=0$). 

\section{Conclusion}

\noindent We have obtained expressions for the mean value of the external
classical force operator for a 1D Schr\"{o}dinger, Klein-Fock-Gordon,
and Dirac particle, in the finite step potential. As we have seen,
only in the Klein-Fock-Gordon theory this quantity cannot be calculated
from the corresponding formula of the mean value. This is because
the corresponding probability density is a discontinuous function
at the point where the step potential is discontinuous ($x=0$). However,
this quantity can always be calculated from the respective wave equation,
which in the latter case is the Feshbach-Villars equation. As we have
seen, the result obtained is proportional to the size of the discontinuity
of the probability density {[}the difference $\varrho_{\mathrm{KFG}}(0+,t)-\varrho_{\mathrm{KFG}}(0-,t)${]},
something that is not expected. In contrast, in the Schr\"{o}dinger and
Dirac theories, the result obtained is simply proportional to the
value that the respective probability density takes at $x=0$, i.e.,
this result is proportional to the average of the (non-existent) discontinuity
of the probability density {[}the sum $\varrho_{\mathrm{A}}(0+,t)+\varrho_{\mathrm{A}}(0-,t)=2\varrho_{\mathrm{A}}(0,t)$,
with $\mathrm{A}=\mathrm{S},\mathrm{D}${]}. 

As mentioned throughout the paper, the external classical force operator
usually appears when studying Ehrenfest's theorem. Here, the most
interesting case is when the potential has finite and/or infinite
discontinuities. We have seen various studies on this subject in Schr\"{o}dinger's
theory (see, for example, Refs. \cite{RefN,RefO,RefP,RefQ} and references
therein), and in Dirac's theory \cite{RefE}. However, in the Klein-Fock-Gordon
theory, as far as we know, there are not many studies that involve
the mean value of this operator. In fact, the main result of the present
article, that is, the result given in Eq. (26), does not seem to have
been published before. We believe that our study may be attractive
to all those interested in fundamental mathematical aspects of quantum
mechanics. 
\begin{acknowledgments}
\noindent The author would like to thank the anonymous referee for
carefully reading the manuscript and for giving such constructive
comments an suggestions.
\end{acknowledgments}

\section{Appendix A}

\noindent We want to prove that Eq. (7) can give the average value
of the external classical force operator even when $V_{0}\rightarrow\infty$,
i.e., for a Schr\"{o}dinger particle in an infinite step potential ($\phi(0-)=0$
and $\phi(0+)=\infty$). In effect, in the latter case we have that
$\Psi(0+,t)=\Psi_{x}(0+,t)=0$ (this is the Dirichlet boundary condition
imposed on $\Psi$ and $\Psi_{x}$ at $x=0+$), and therefore, $\varrho_{\mathrm{S}}(0+,t)=0$,
and although we have that $\phi(0+)=\infty$, we also have that $\phi(0+)\Psi(0+,t)=\infty\times0=0$
(below, we formally prove this last result, but it must be true because
the Schr\"{o}dinger equation must also be verified at $x=0+$); thus,
from Eq. (7), we obtain 
\[
0=+\frac{\hbar^{2}}{2\mathrm{m}}|\Psi_{x}(0-,t)|^{2}+\langle\hat{f}\rangle_{\mathrm{S}}\quad\Rightarrow\quad\langle\hat{f}\rangle_{\mathrm{S}}=-\frac{\hbar^{2}}{2\mathrm{m}}|\Psi_{x}(0-,t)|^{2}.\tag{A1}
\]
Incidentally, in Ref. \cite{RefR}, the latter result was obtained
directly from the same procedure presented when deriving Eq. (7),
assuming from the beginning that $V_{0}\rightarrow\infty$. On the
other hand, in Ref. \cite{RefN}, the result in Eq. (A1) was obtained
directly from the Ehrenfest theorem for a (free) Schr\"{o}dinger particle
on a half line (in our case, the region $x\in(-\infty,0]$). Finally,
let us now prove that the product $\phi\Psi$ to the right of the
point $x=0$ where the potential makes an infinite jump (i.e., at
the extensive impenetrable barrier located in the region $x>0$),
vanishes, i.e., $\phi(0+)\Psi(0+,t)=0\,(=\infty\times0)$. In effect,
integrating the Schr\"{o}dinger equation (Eq. (2) with $\mathrm{A}=\mathrm{S}$)
from $x=0-$ to $x=0+$ yields
\[
0=\frac{\hbar^{2}}{2\mathrm{m}}\Psi_{x}(0-,t)+\int_{0-}^{0+}\mathrm{d}x\,\phi(x)\Psi(x,t),\tag{A2}
\]
where we have made use of the condition $\Psi_{x}(0+,t)=0$. Clearly,
the relation in Eq. (A2) is satisfied by writing the following relation:
\[
\phi(x)\Psi(x,t)=-\frac{\hbar^{2}}{2\mathrm{m}}\Psi_{x}(0-,t)\delta(x)\tag{A3}
\]
(this is because the Dirac delta function satisfies $\int_{0-}^{0+}\mathrm{d}x\,\delta(x)=1$).
Thus, from Eq. (A3), we obtain the desired result,
\[
\phi(0+)\Psi(0+,t)=0,\tag{A4}
\]
and also the (expected) result $\phi(0-)\Psi(0-,t)=0$ (remember that
$\delta(0+)=\delta(0-)=0$), however, $\phi(0)\Psi(0,t)=\infty$ (this
is because $\delta(0)=\infty$). Incidentally, the result given by
Eq. (A3) was also obtained and discussed in Ref. \cite{RefS}. 

\section{Appendix B}

\noindent We want to prove that the boundary conditions for $\Psi$
and $\Psi_{x}$ given in Eqs. (21) and (22) arise directly from the
Feshbach-Villars equation. In effect, integrating the latter equation
(Eq. (2) with $\mathrm{A}=\mathrm{KFG}$) from $x=0-$ to $x=0+$
gives
\[
0=-\frac{\hbar^{2}}{2\mathrm{m}}\left.\left[\,(\hat{\tau}_{3}+\mathrm{i}\hat{\tau}_{2})\Psi_{x}(x,t)\,\right]\right|_{0-}^{0+}+0+0,\tag{B1}
\]
from which the boundary condition given in Eq. (22) is obtained. Likewise,
integrating the Feshbach-Villars equation first from $x=-\ell$ (with
$\ell>0$) to $x$ and then once more from $x=0-$ to $x=0+$ gives
\[
\mathrm{i}\hbar\int_{0-}^{0+}\mathrm{d}x\int_{-\ell}^{x}\mathrm{d}y\,\dot{\Psi}(y,t)=-\frac{\hbar^{2}}{2\mathrm{m}}\left.\left[\,(\hat{\tau}_{3}+\mathrm{i}\hat{\tau}_{2})\Psi(x,t)\,\right]\right|_{0-}^{0+}+\frac{\hbar^{2}}{2\mathrm{m}}(\hat{\tau}_{3}+\mathrm{i}\hat{\tau}_{2})\Psi_{x}(-\ell,t)\int_{0-}^{0+}\mathrm{d}x
\]
\[
+\mathrm{m}c^{2}\int_{0-}^{0+}\mathrm{d}x\int_{-\ell}^{x}\mathrm{d}y\,\hat{\tau}_{3}\Psi(y,t)+\int_{0-}^{0+}\mathrm{d}x\int_{-\ell}^{x}\mathrm{d}y\,\phi(y)\Psi(y,t),\tag{B2}
\]
that is, 
\[
0=-\frac{\hbar^{2}}{2\mathrm{m}}\left.\left[\,(\hat{\tau}_{3}+\mathrm{i}\hat{\tau}_{2})\Psi(x,t)\,\right]\right|_{0-}^{0+}+0+0+0,\tag{B3}
\]
from which the boundary condition given in Eq. (21) is obtained. Note
that for any discontinuous potential at $x=0$, and not only for the
jump potential, we obtain the continuity of $(\hat{\tau}_{3}+\mathrm{i}\hat{\tau}_{2})\Psi$
and $(\hat{\tau}_{3}+\mathrm{i}\hat{\tau}_{2})\Psi_{x}$ at that point.

\section*{Note}

\noindent This paper will be published in the Revista Brasileira de Ensino de F\'{\i}sica (RBEF) in 2021.

\end{document}